\documentclass[aps,prm,reprint,twocolumn,superscriptaddress,showpacs,amssymb]{revtex4-1}
\usepackage[english]{babel}
\usepackage{graphics}
\usepackage{graphicx}
\usepackage{epsfig}
\usepackage{amssymb}
\usepackage{dcolumn}
\usepackage{bm}
\usepackage{color}
\usepackage{natbib}
\usepackage{lineno}
\usepackage{sidecap}
\usepackage{verbatim}  
\usepackage{vector}  
\usepackage{wrapfig}
\usepackage{enumerate}
\usepackage{sidecap}
\usepackage{amsmath}
\usepackage{hyperref}

\begin{document}

\title{Massive Dirac fermions in layered BaZnBi$_2$}

\author{S.\ Thirupathaiah}
\email{setti@bose.res.in}
\affiliation{Leibniz Institute for Solid State Research, IFW Dresden, D-01171 Dresden, Germany.}
\affiliation{S. N. Bose National Center for Basic Sciences, JD Block, Sector III, Salt Lake City, Kolkata, 700106, India.}
\author{D. Efremov}
\affiliation{Leibniz Institute for Solid State Research, IFW Dresden, D-01171 Dresden, Germany.}
\author{Y.\ Kushnirenko}
\affiliation{Leibniz Institute for Solid State Research, IFW Dresden, D-01171 Dresden, Germany.}
\author{E.\ Haubold}
\affiliation{Leibniz Institute for Solid State Research, IFW Dresden, D-01171 Dresden, Germany.}
\author{T. K.\ Kim}
\affiliation{Diamond Light Source, Harwell Campus, Didcot, OX11 0DE, UK.}
\author{B. R. Pienning}
\affiliation{Leibniz Institute for Solid State Research, IFW Dresden, D-01171 Dresden, Germany.}
\author{I. \ Morozov}
\affiliation{Leibniz Institute for Solid State Research, IFW Dresden, D-01171 Dresden, Germany.}
\affiliation{Lebedev Physical Institute, Russian Academy of Sciences, 119991 Moscow, Russia.}
\affiliation{Lomonosov Moscow State University, Moscow, 119991, Russian Federation.}

\author{S.\ Aswartham}
\affiliation{Leibniz Institute for Solid State Research, IFW Dresden, D-01171 Dresden, Germany.}
\author{B. B\"uchner}
\affiliation{Leibniz Institute for Solid State Research, IFW Dresden, D-01171 Dresden, Germany.}
\author{S. V.\ Borisenko}
\email{s.borisenko@ifw-dresden.de}
\affiliation{Leibniz Institute for Solid State Research, IFW Dresden, D-01171 Dresden, Germany.}

\date{\today}

\begin{abstract}
    Using angle-resolved photoemission spectroscopy (ARPES) and density functional theory (DFT) we study the electronic structure of layered BaZnBi$_2$. Our experimental results show no evidence of  Dirac states in BaZnBi$_2$ originated either from the bulk or the surface. The calculated band structure without spin-orbit interaction shows several linear dispersive band crossing points throughout the Brillouin zone. However, as soon as the spin-orbit interaction is turned on,  the band crossing points are significantly gapped out.  The experimental observations are in good agreement with our DFT calculations. These observations suggest that the Dirac fermions in  BaZnBi$_2$ are trivial and massive.   We also observe experimentally that the electronic structure of BaZnBi$_2$ comprises of several linear dispersive bands in the vicinity of Fermi level dispersing to a wider range of binding energy.

\end{abstract}

\maketitle

\subsection{INTRODUCTION}

Discovery of topological insulators with Dirac fermions at the surface~\cite{Hasan2010} initiated an intense search of topological phases having Dirac~\cite{Young2012, Wang2012a, Wang2013a,  Borisenko2014, Neupane2014a, Liu2014a, Thirupathaiah2018a} and Weyl fermions in the bulk~\cite{Huang2015,Zhang2015a,Ghimire2015,Shekhar2015, Xu2016, Tamai2016, Deng2016, Huang2016, Liang2016,Wang2016a, Jiang2017, Thirupathaiah2017, Thirupathaiah2017a}. All these systems have potential applications in topological quantum computations~\cite{Nayak2008},  spintronics~\cite{Datta1990, Ziutifmmodecuteclseci2004} and topotronics~\cite{Hesjedal2017}. The Dirac fermions emerging in three dimensional  band crossings are usually protected by both the time-reversal and crystal symmetries. These bulk Dirac fermions are characterized by a four-fold degeneracy at such crossing point~\cite{Burkov2011, Young2012, Wang2012a}. On the other hand, the Weyl fermions, being formed due to breaking of either time-reversal or inversion crystal symmetry, are characterized by a two-fold degeneracy\cite{Wan2011, Borisenko2015}. When symmetry protection of the band crossings disappears,  a gap opens and the system becomes topologically trivial.

Materials of the type AMnB$_2$ (A=Ca, Sr, Ba, Eu, Yb; B=Bi and Sb) are widely known for possessing the Dirac fermions near the Fermi level~\cite{Park2011a,  Wang2011b, Wang2011a, Wang2012c, Lee2013, Farhan2014, Feng2014, Guo2014, Jo2014, May2014, Jia2014, Borisenko2015, Zhang2016, Li2016, Liu2016, He2017}. Origin of these Dirac states was understand by considering the main structural constituent of these materials - a square net of Bi atoms. The strong hybridization results in a large bandwidth with large portions of linear dispersions. If there is a reason for folding of such electronic structure, e.g. doubling of the unit cell in real space, the linear dispersions start to cross resulting in a multitude of Dirac crossings in the Brillouin zone~\cite{Borisenko2015}. However, in many cases such crossings are gapped out by the spin-orbit interaction~\cite{Park2011a,  Lee2013,  Feng2014, Guo2014, Li2016, Masuda2016, Liu2016, Wang2016d, Liu2017a}. In some materials the non-trivial phases can still be observed despite the gapped Dirac states~\cite{Wang2016e,Wang2016e, Huang2017}.

Furthermore, these compounds show unsaturated linear magnetoresistance under the external magnetic fields  which are attributed to the Dirac fermions~\cite{He2012, Wang2016c} or to the proximity of long range magnetic ordering~\cite{Zhang2016, Kormondy2018}. Interestingly, isostructural  to AMnBi$_2$,  BaZnBi$_2$ is a nonmagnetic compound but yet shows linear magnetoresistance~\cite{Wang2017}. Despite the property of linear magnetoresistance in BaZnBi$_2$, the presence of Dirac states near the Fermi level in this system is not yet established.
While some studies do not find evidence of the Dirac states ~\cite{Wang2017, Ren2018},  the other ones report the existence Dirac states ~\cite{Zhao2018} in this system. Since BaZnBi$_2$ is a nonmagnetic system and yet shows the linear magnetoresistance, it is an ideal system to examine the relation between the linear magnetoresistance, the magnetism, and the possible existence of the Dirac states.

\begin{figure}[htbp]
	\centering
		\includegraphics[width=0.49\textwidth]{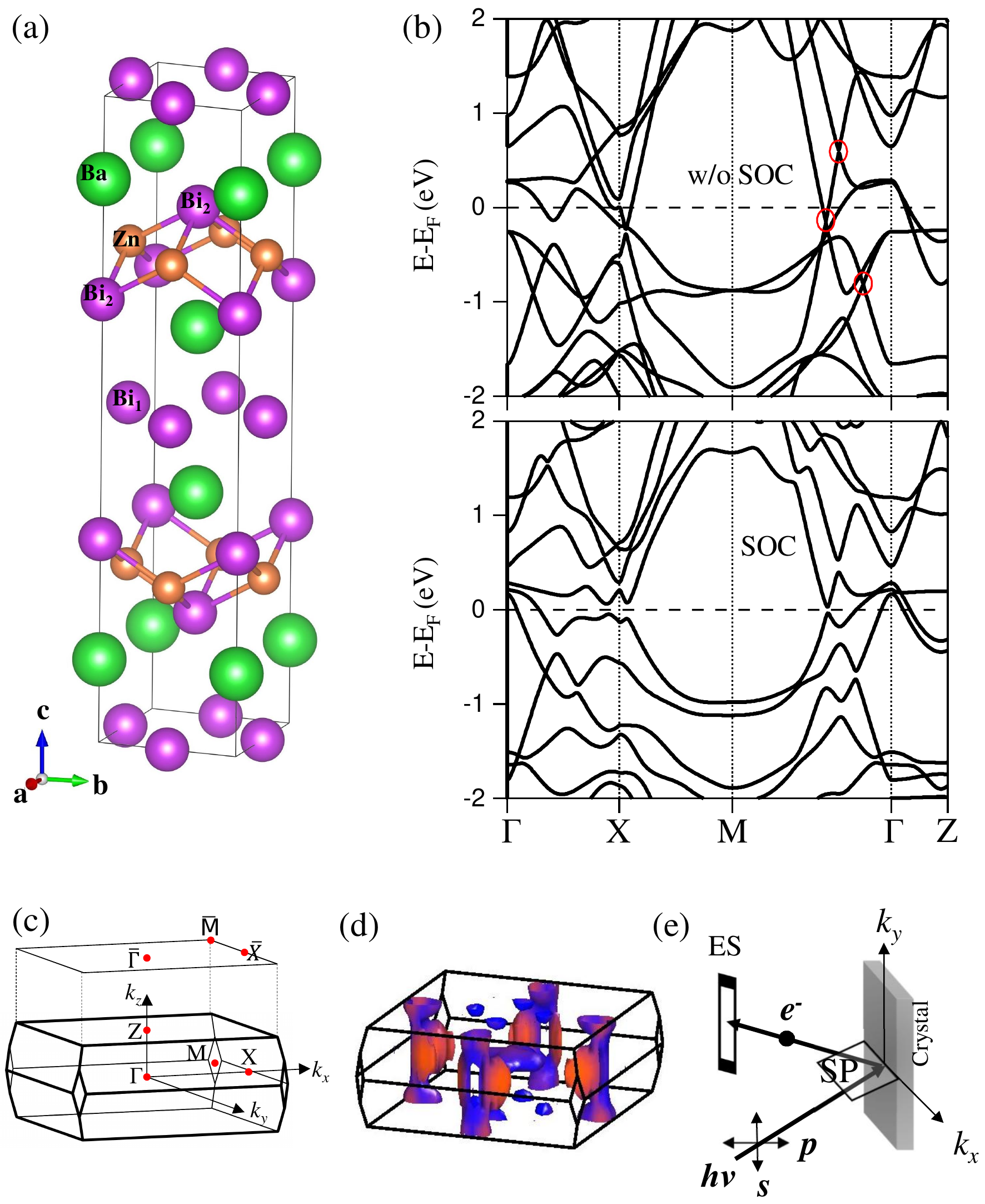}
	\caption{(a) Tetragonal crystal structure of BaZnBi$_2$. (b) Calculated bulk band structure without SOC (top panel) and with SOC (bottom panel). (c) 3D view of the Brillouin zone with projected surface 2D Brillouin zone on the top. (d) 3D Fermi surface obtained with SOC. (e) Experimental measuring geometry in which the $s$ and $p$ polarized lights are defined with respect to the scattering plane (SP) and the analyzed entrance slit (ES).}
	\label{1}
\end{figure}


\begin{figure*}[htbp]
	\centering
		\includegraphics[width=0.85\textwidth]{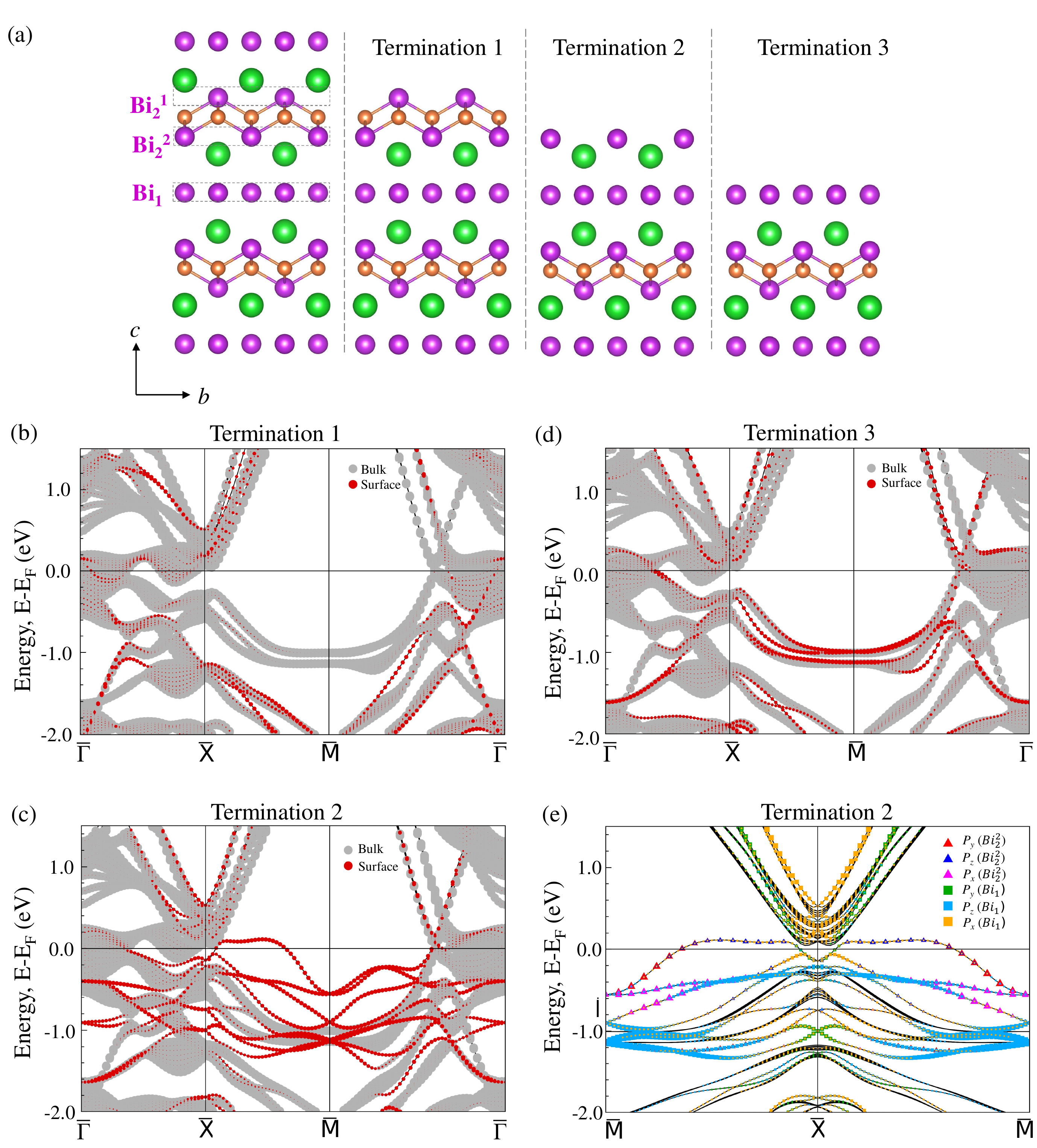}
	\caption{Slab calculations performed on BaZnBi$_2$ including SOC. (a) Shows the supercell of BaZnBi$_2$ in the $b-c$ plane (left panel),
termination1 (T1) leaves out the Bi$_{2}^{1}$ layer on the top of the sample surface (middle panel), the termination2 (T2) leaves out the Bi$_{2}^{2}$ layer on the top of the sample surface (middle panel), and the termination3 (T3) leaves out the Bi$_{1}$ layer on the top of the sample surface (right panel). (b) Surface sates obtained after T1 cleavage. (c) Surface sates obtained after T2 cleavage. (d) Surface sates obtained after T3 cleavage. (d) Orbital resolved surface states are shown in the $\overline{X}$-$\overline{M}$ high symmetry line after T2 cleavage.}
	\label{2}
\end{figure*}

In this paper, we report the low-energy electronic structure of BaZnBi$_2$ studied using the high-resolution angle-resolved photoemission spectroscopy and the density functional theory calculations. Our experimental results show no evidence of bulk or surface Dirac states near the Fermi level in this system. However, we do observe several linear dispersive band crossing points near the Fermi level when calculated without including the spin-orbit interaction. On the other hand, these band crossing points are gapped out as soon as the spin-orbit interaction is turned on.  Thus, our results suggest that the Dirac states in BaZnBi$_2$ are trivial and massive, analogous to the Dirac states in graphene~\cite{Novoselov2005, Brey2006, Varykhalov2008}.   Nevertheless, despite the gap opening at the nodes, there exist several linear dispersive bands crossing the Fermi level as seen from our experimental data.


\subsection{EXPERIMENTAL DETAILS}

Single crystals of BaZnBi$_2$ were synthesized by the self-flux method. The elements of Ba, Zn and Bi were mixed in the composition of BaZnBi$_6$ and then was placed in an alumina crucible, which in turn was
sealed inside the evacuated quartz tube. The whole setup was heated to 1000$^\circ$C and
held at that temperature for 10 hours, and then slowly cooled to 400$^\circ$C  at rate
of 2$^\circ$C/h. Excess flux was removed by centrifugation at this
temperature. The remaining Bi flux residue was removed by cleaving the crystal.  ARPES measurements were performed in the Diamond light source at I05 beamline which is equipped with SCIENTA R4000 analyzer. During the measurements the sample temperature was kept at 5 K.  The energy resolution was set between 10 and 20 meV depending on the incident photon energy and the angle resolution was set at 0.3$^\circ$.

 \subsection{BAND STRUCTURE CALCULATIONS}

Bulk band structure calculations were performed  using density functional theory (DFT) within the generalized gradient approximation (GGA) of Perdew, Burke and Ernzerhof (PBE) exchange and correlation potential~\cite{Perdew1996} as implemented in the Quantum Espresso simulation package~\cite{QE-2009}. Norm conserving scalar relativistic and fully relativistic pseudopotentials were used to perform the calculations without spin-orbit coupling (SOC) and with SOC, respectively. The electronic wave function is expanded using the plane waves up to a cutoff energy of 50 Ry (680 eV). Brillouin zone sampling was done over a 24 $\times$ 24 $\times$ 10 Monkhorst-Pack k-grid. The slab calculations are performed using the DFT within the fully relativistic GGA approximation as implemented in the Full Potential Local Orbital band structure package (FPLO)~\cite{Koepernik1999}. To obtain the surface states, we projected the Bloch wave functions onto the atomic-like Wannier functions, and constructed the tight-binding model Hamiltonian. Then the tight-binding model was mapped onto a slab geometry.

\begin{figure} [htbp]
	\centering
		\includegraphics[width=0.49\textwidth]{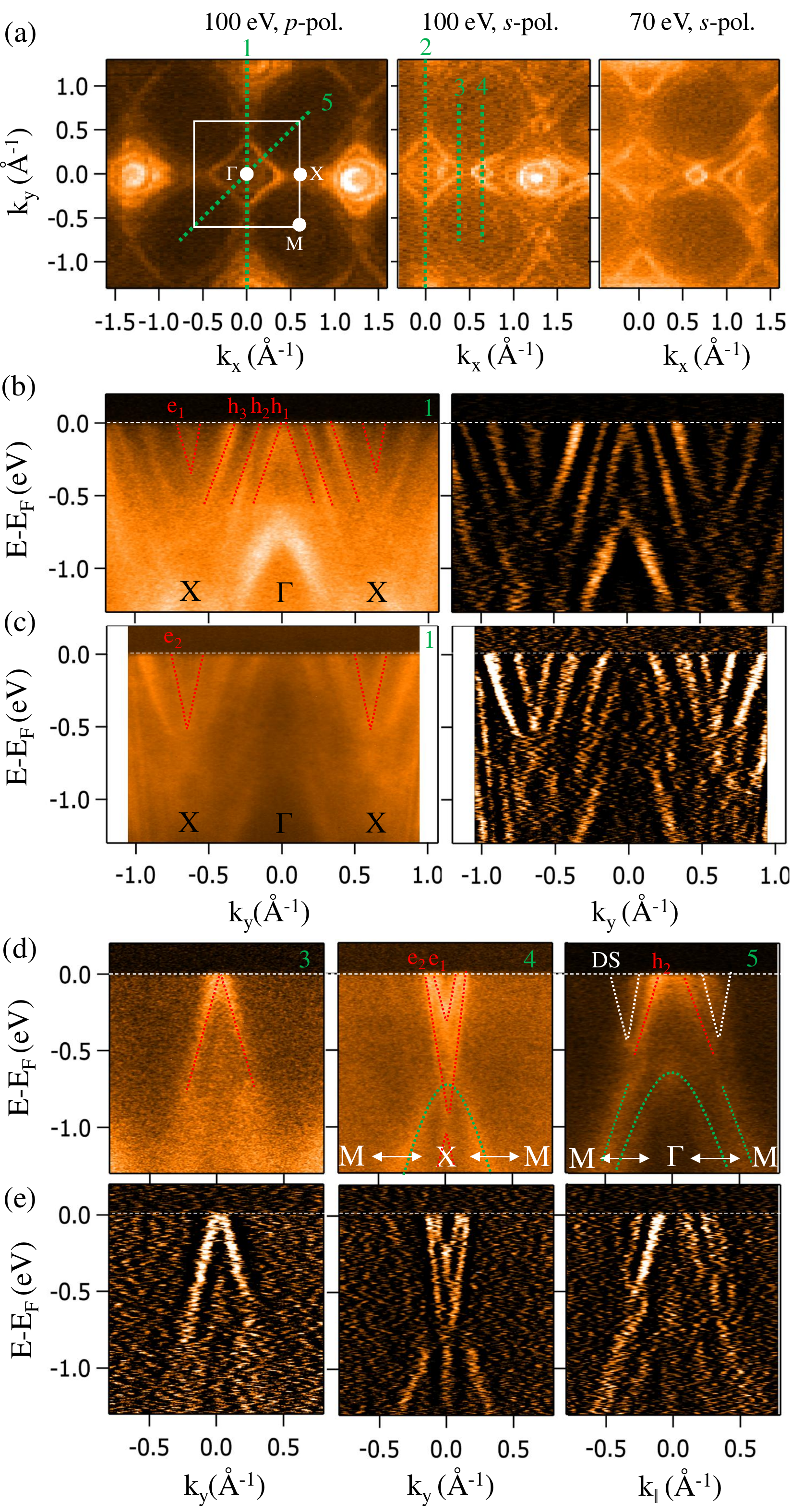}
	\caption{ARPES data of BaZnBi$_2$. (a) Fermi surface maps taken with different photon energies (70 and 100 eV) and polarizations ($s$ and $p$). (b) and (c) are the energy distribution maps (EDM) taken along the cuts~1 and 2 (left panel) and corresponding second derivative (right panel) using $p$ and $s$ polarized photons,  respectively, along the $\Gamma - X$ high symmetry line. (d) Shows EDMs taken along the cuts~3, 4, and 5 from left to right, respectively. (e) Shows respective second derivatives of (d).}
	\label{3}
\end{figure}

 \subsection{RESULTS AND DISCUSSIONS}

Figure ~\ref{1} shows the calculated bulk band structure of BaZnBi$_2$.  Top panel in Fig.~\ref{1} (b) shows the electronic structure calculated without spin-orbit interaction and the bottom panel in Fig.~\ref{1} (b) shows the electronic structure calculated  with SOC. Fig.~\ref{1} (d) depicts the 3D view of the Fermi surface obtained under SOC. As can be noticed in Fig.~\ref{1} (b), there exist several band crossing points (Dirac points) in the vicinity of the Fermi level as highlighted by the red circles in the electronic structure calculated without SOC. However,  all of them are gapped out as soon as the SOC is turned on in the calculations as shown in the bottom panel of Fig.~\ref{1} (b). Important to notice here that our DFT calculations show no bulk Dirac states at the $X$ point. We will further discuss this point at a later stage.

To further examine whether there exists any Dirac fermions originated from the surface, we performed slab calculations on two different top Bi layers as demonstrated in Fig.~\ref{2} (a). In this compound it is possible to have different Bi layers on the sample surface with different cleavages. As shown in Fig.~\ref{2} (a), termination1 (T1) leaves out the Bi$_{2}^{1}$ layer on top of the sample surface that is bonded with the bottom Zn layer,  the termination2 (T2) leaves out the Bi$_{2}^{2}$ layer on  top of the sample surface after breaking the bonds with Zn layer, and the termination3 (T3) leaves out the Bi$_{1}$ layer on  top of the sample surface which is an isolated layer. Thus,  the top Bi surface layers produced with T1, T2, and T3 cleavages are different from each other. Specifically, the T2 cleavage produces a polar surface  due to Bi-Zn bond breaking. The other two cleavages, T1 and T2, produce neutral surface. Since during the experiment getting one of these three surface Bi layers has equal chances, we performed slab calculations for all Bi$_{2}^{1}$,  Bi$_{2}^{2}$, and Bi$_{1}$ surface layers.  Corresponding slab calculations are shown in   Figs.~\ref{2} (b),(c), and (d) after T1, T2, and T3 cleavages, respectively. Here black colored band structure represents the bulk and the red colored band structure represents the surface. Fig.~\ref{2} (e) shows the orbital resolved surface states with T2 cleavage. From Figs.~\ref{2} (b)-(d) it is evident that the different terminations (T1, T2, \& T3) lead to an entirely different set of surface states, more specifically, the difference in the surface states is substantial near the $\overline{M}$ point. Furthermore, one can notice from Fig.~\ref{2} (d) that all the surface states near the Fermi level are mainly contributed by the Bi-2$p$ orbital characters. Most importantly, our slab calculations also do not predict any noticeable surface Dirac states at the $X$ point.

\begin{figure} [htbp]
	\centering
		\includegraphics[width=0.49\textwidth]{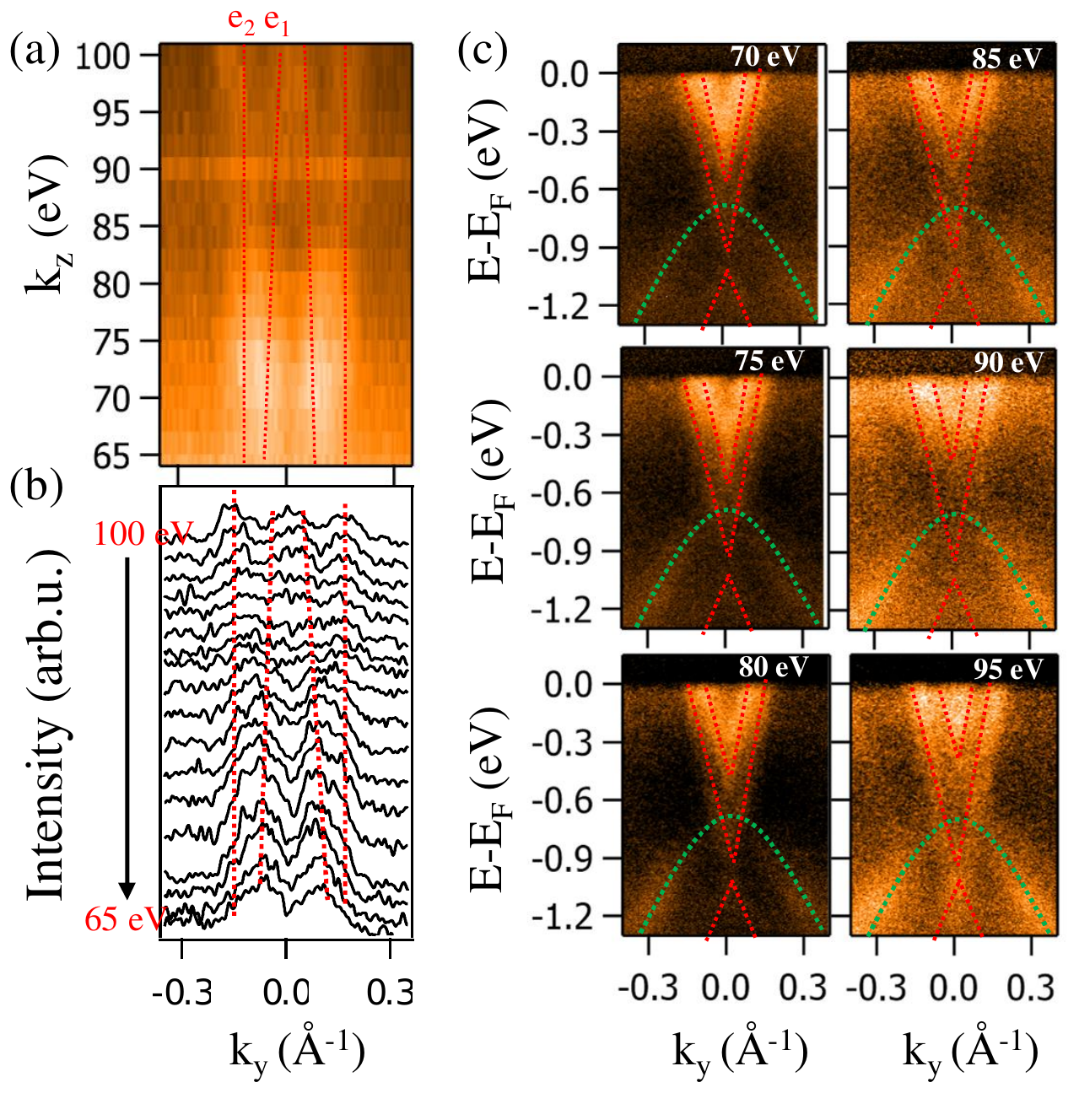}
	\caption{ (a) Fermi surface map taken along the $X-M$ orientation in the $k_y-k_z$ plane. (b) Photon energy dependent momentum distribution curves (MDCc) extracted from (a) around the Fermi level with an energy integration of 15 meV .  (c) Photon energy dependent EDMs. In the figure, the red dashed lines represent linear dispersive bands, while the green dashed curves represent the quadratic bands.}
	\label{4}
\end{figure}

Figure ~\ref{3} shows the ARPES data of BaZnBi$_2$. Fig.~\ref{3}(a) depicts Fermi surface maps  measured with a photon energy of 100 eV using $p$-polarized light (left panel)and $s$-polarized light (middle panel). Similarly, the right most panel in Fig.~\ref{3}(a) shows the FS map measured with a photon energy of 70 eV using $s$-polarized light. Energy distribution maps (EDMs) measured along the cuts 1 and 2 [as shown on the FS maps in  Fig.~\ref{3} (a)] and their corresponding second derivatives are shown in Fig.~\ref{3} (b) and (c), respectively. Similarly, the EDMs along the cuts 3-5 and their corresponding second derivatives are shown in Fig.~\ref{3}(d). As can be seen from Fig.~\ref{3}(a), the Fermi surface consists of several square shaped outer Fermi sheets and circle shaped inner Fermi sheets around the $\Gamma$ point.   From the EDM cut taken along $\Gamma-X$ orientation as shown in Fig.~\ref{3}(b) and (c), we could resolve three hole-like, $h_1$, $h_2$ and $h_3$,  linear band dispersions crossing the Fermi level at the $\Gamma$ point and two electron-like, $e_1$ and $e_2$,  linear dispersions crossing the Fermi level at the $X$ point. Similarly, from the EDM cut 3 [left panel in Fig.~\ref{3} (d)], we could resolve one linear dispersive hole-like band, $h_3$,  whose band top is just touching the Fermi level. From the EDM cut taken along the $X-M$ orientation [middle panel in Fig.~\ref{3} (d)] two electron-like,$e_1$ and $e_2$,  linear band dispersions are noticed at the $X$-point. From the EDM cut taken along the $\Gamma-M$ orientation [right most panel in Fig.~\ref{3} (d)] we could observe the upper cone of the gapped Dirac states as shown by the white dashed lines. As can be seen in the figure [right most panel in Fig.~\ref{3} (d)], it is very clear that there is a gap in the place of lower part of the Dirac cone, consistent with our DFT calculations performed with SOC.  We further estimated the mass enhancement factor of the upper cone of the gapped Dirac states to $\frac{m_{SO}^{*}}{m}$ = 1.4 by using our calculated band structure. Thus, the Dirac states are acquiring mass under the spin-orbit interaction.


Although experimentally we did not find any evidence for the Dirac states in this system, Ref.~\onlinecite{Zhao2018} showed the Dirac states near the $X$ point and as well along the $\Gamma-M$ orientation in their ARPES data. In order to verify this more carefully and to further elucidate nature of the states near the $X$ point, we performed photon energy dependent measurements in the range of 65 eV to 100 eV taken in steps of 3 eV using $p$ polarized light along the $\Gamma-X$ high symmetry line as shown in Fig.~\ref{4}. Fig.~\ref{4} (a) depicts the Fermi surface map measured in the $k_y-k_z$ plane. From Fig.~\ref{4} (a) we can resolve two Fermi sheets shown by the red dashed lines on the FS map. Photon energy dependent momentum distribution curves are shown in Fig.~\ref{4} (b). To further elucidate the $k_z$ dependent band structure at the $X$-point we showed EDMs as function of photon energy in Fig.~\ref{4} (c). From Figs.~\ref{4} (a)-(c) it is clear that the outer-electron pocket shows no $k_z$ dispersion,  while the inner-electron pocket shows a significant change in the momentum vector in going from 65 eV to 100 eV photon energy. Our careful analysis of the band structure as a function of photon energy shows no evidence of Dirac states in the $k_z$ direction at the $X$ point. Comparing more rigorously our experimental data with that of Ref.~\onlinecite{Zhao2018} to resolve the discrepancies on the presence of Dirac states near the $X$ point, in Ref.~\onlinecite{Zhao2018} the Dirac states are concluded based on finding one upper and one lower band with a speculated Dirac node at around 200 meV. However,  from our high resolution ARPES data we clearly see two linearly dispersive upper states [see Figs.~\ref{3}(d), (d) and Fig.~\ref{4}], $e_1$ and $e_2$, whose band bottoms are at approximately 200 meV and 900 meV below the Fermi level,  respectively. Further,  we found two lower bands whose band tops are at 700 meV and 1 eV, respectively. Importantly, one of the two lower bands whose band top is at 700 meV is a parabolic band. Thus, our experimental data completely rule out the existence of Dirac states near the $X$ point in this system. Also it is worth mentioning here that, neither our DFT calculations did predict them.  We further noticed clearly that the gapped out Dirac states from the EDM measured along the $\Gamma-M$ direction [see white dashed lines in the right most panel of Fig.~\ref{3}(d)], again in contrast to the observations made in Ref.~\onlinecite{Zhao2018} we did not find any clear evidence on the presence of Dirac states. Thus, our experimental observation on the absence of gapless Dirac states in this compound is in very good agreement with previous reports on this system~\cite{Wang2017, Ren2018}.


As we systematically demonstrated above, our experimental results show no evidence of gapless Dirac states in this compound originated either from the bulk or the surface. However, we noticed several linear dispersive bands crossing the Fermi level near the  $\Gamma$ and $X$ high symmetry points. This observation is also consistent with the previous reports on  AMnBi$_2$ type compounds~\cite{Feng2014, Borisenko2015, Kealhofer2018}, in which the linear dispersive bands crossing the Fermi level have been reported by both the ARPES studies and DFT calculations.  Moreover, the Dirac states are gapped out in AMnBi$_2$ type compounds under the spin-orbit interactions, much like the case in BaZnBi$_2$~\cite{Park2011a, Farhan2014, Feng2014}.
This large overlapping of the band structure between  BaZnBi$_2$ and AMnBi$_2$ are suggesting that both share a common mechanism for the presence of Dirac states as predicted from the DFT calculations, that is the Bi square net present in both systems. At the same time,  these Dirac states are gapped out at the Dirac node in both compounds  with the spin-orbit interactions. Thus, in these magnetic AMnBi$_2$ and nonmagnetic BaZnBi$_2$,   the electronic structure is largely governed by the crystal structure rather than by the magnetic ordering although there is non-negligible magnetic effect on the band structure~\cite{May2014, Guo2014, Zhang2016}.

Next,  BaZnBi$_2$ is known to show a large linear magnetoresistance under the external magnetic fields~\cite{Wang2017, Ren2018, Zhao2018}. This property is attributed to the electron-hole compensation~\cite{Wang2017} and the linear dispersive Dirac states~\cite{Zhao2018}. The charge compensation theory is widely applied for understanding the quadratic field dependent magnetoresistance in the bulk crystals~\cite{Ali2014}.  On the other hand, from our experimental data we found no evidence of massless Dirac states in this compound. Hence, the Dirac states can not be the reason behind the observed linear magnetoresistance in this compound. As noticed from our experimental data and well from the slab calculations, there exist several linear dispersive bulk and surface states near the Fermi level dispersing to a wider range of binding energy (see Figs.~\ref{3} and ~\ref{4}). Perhaps, these linear dispersive bulk and surface states in the vicinity of the Fermi level could be a reason behind the observed linear magnetoresistance~\cite{Thirupathaiah2018, Abrikosov1998} in this compound. The same explanation maybe is true also for the recorded linear magnetoresistance in the AMnBi$_2$ type compounds~\cite{Wang2011a, Wang2012c, Li2016, Wang2016d}.

In conclusion, we examined the low-energy electronic structure of BaZnBi$_2$ by means of ARPES and DFT calculations.  Our experimental results show no evidence bulk or surface massless Dirac states near the Fermi level in this system. However, we do notice several linear dispersive bands crossing the Fermi level at both the $\Gamma$ and $X$ high symmetry points. The bulk band structure obtained with DFT calculations without SOC shows several linear dispersive Dirac states. However, all these Dirac states are gapped out and acquire a significant mass as soon as the SOC is turned on. Thus, our results suggest that the Dirac states predicted in this system are trivial and massive. Since experimentally we could not find the massless Dirac states in this compound,  the observed linear magnetoresistance may have a different origin rather than the proposed Dirac states.

This work was supported under DFG Grant No. BO 1912/7-1. S.T. acknowledges support by the Department of Science and Technology, India through the INSPIRE-Faculty program (Grant No. IFA14 PH-86). The authors thank K. K\"opernik for useful discussions and U. Nitzsche for technical support. The authors also thank G. Shipunov and S. M\"uller-Litvanyi.  D.E. and I.M. acknowledge the support by RSCF and DFG through the grant RSF-DFG 16-42-01100.  We acknowledge the Diamond Light Source for the time on Beamline I05 under the Proposal SI18586-1.

\bibliography{BaZnBi2}

\end{document}